\documentclass[%
 reprint,
 amsmath,amssymb,
 aps,
]{revtex4-2}
\usepackage{graphicx}
\usepackage{ textcomp}
\usepackage{amssymb}
\usepackage[colorlinks,linkcolor=blue,anchorcolor=blue,citecolor=blue,urlcolor=black]{hyperref}
\usepackage[utf8]{inputenc}
\usepackage{amsmath}
\usepackage{tipa}
\usepackage{comment}
\usepackage{mathtools}
\usepackage{braket}
\usepackage{listings}
\usepackage{xcolor}
\usepackage{soul}
\definecolor{Darkgreen}{rgb}{0,0.4,0}
\definecolor{listinggray}{gray}{0.9}
\definecolor{lbcolor}{rgb}{0.9,0.9,0.9}
\begin{document}

\title{Observation of bosonic stimulation in light scattering
}

\author{Yu-Kun Lu}
\thanks{Corresponding author. E-mail: yukunlu@mit.edu.}
\affiliation{Research Laboratory of Electronics, MIT-Harvard Center for Ultracold Atoms, and Department of Physics, Massachusetts Institute of Technology, Cambridge, Massachusetts 02139, USA}

\author{Yair Margalit}
\affiliation{Research Laboratory of Electronics, MIT-Harvard Center for Ultracold Atoms, and Department of Physics, Massachusetts Institute of Technology, Cambridge, Massachusetts 02139, USA}

\author{Wolfgang Ketterle}
\affiliation{Research Laboratory of Electronics, MIT-Harvard Center for Ultracold Atoms, and Department of Physics, Massachusetts Institute of Technology, Cambridge, Massachusetts 02139, USA}
 \date{\today}
 \begin{abstract}
For bosons, the transition rate into an already occupied quantum state is enhanced by its occupation number: the effect of bosonic stimulation. 
Bosonic enhancement of light scattering has been predicted more than 30 years ago but has not been observed before. Here we theoretically investigate and experimentally demonstrate this effect. We show that the bosonic enhancement factor for a harmonically trapped gas is bounded by a universal constant $\zeta(2)/\zeta(3)$ above the phase transition to a Bose-Einstein condensate (BEC), and depends linearly on the BEC fraction just below the phase transition. Bosonically enhanced light scattering is observed and characterized above and below the phase transition, and the effect of interactions is discussed. For a multi-level system, bosonic enhancement is reduced because bosonic stimulation occurs only for Rayleigh scattering, but not for Raman scattering. 
 \end{abstract}
\maketitle
Bosonic stimulation occurs for bosonic particles transitioning into a final state with non-zero occupation number $n$: the transition rate is enhanced by a factor of $n+1$. Since most elementary particles are fermions, most observations of bosonic stimulations were related to photons, with the laser as the most dramatic manifestation. The field of ultracold atoms has created new opportunities to realize other paradigmatic phenomena for bosonic stimulation: it was observed in the formation of BEC~\cite{miesner1998bosonic}, superradiance~\cite{inouye1999superradiant}, matter-wave amplification~\cite{inouye1999phase,kozuma1999phase} and enhanced density fluctuations~\cite{PhysRevLett.96.130403,blumkin2013observing}. However, one phenomenon has never been observed, although it was predicted more than 30 years ago: bosonic enhancement of light scattering~\cite{svistunov1990resonance, PhysRevA.43.6444, PhysRevA.49.3799, PhysRevA.51.3896, PhysRevA.52.3033, PhysRevLett.72.2375, PhysRevA.55.1140}. One reason why the experimental observation has been elusive~\cite{footnote, PhysRevLett.116.173602, footnote2,doi:10.1126/science.abh3470} is the requirement of high densities, ideally many atoms per $\lambdabar^3$ ($\lambdabar = \lambda/2\pi$ is the reduced wavelength). At such densities, ultracold atomic clouds become short-lived due to inelastic collisions~\cite{inguscio1999bose,top2021spin}. For the same reason, the fermionic counterpart, Pauli blocking of light scattering, was demonstrated only very recently~\cite{doi:10.1126/science.abi6153,doi:10.1126/science.abh3470,doi:10.1126/science.abh3483}. 

In this work, we theoretically investigate and experimentally observe bosonic enhancement of light scattering in an ultracold Bose gas of sodium atoms. We obtain new analytic results of the enhancement factor for non-interacting bosons in several regimes. Experimentally, we prepared a dense cloud of bosons with a peak density up to $2\times10^{15}\text{cm}^{-3}$ and a BEC transition temperature of $T_c\approx10\mu K$, achieving the highest $T_c$ in ultracold atoms except for atomic hydrogen~\cite{PhysRevLett.81.3811} and the highest critical density. The bosonic enhancement of light scattering was observed in two different regimes: for scattering within the thermal cloud and for scattering between the BEC and thermally occupied states. We show how interactions modify bosonic stimulation by modifying the overlap between condensate and thermal cloud, and by changing the pair correlation function.


\begin{figure}
\includegraphics[width=\columnwidth]{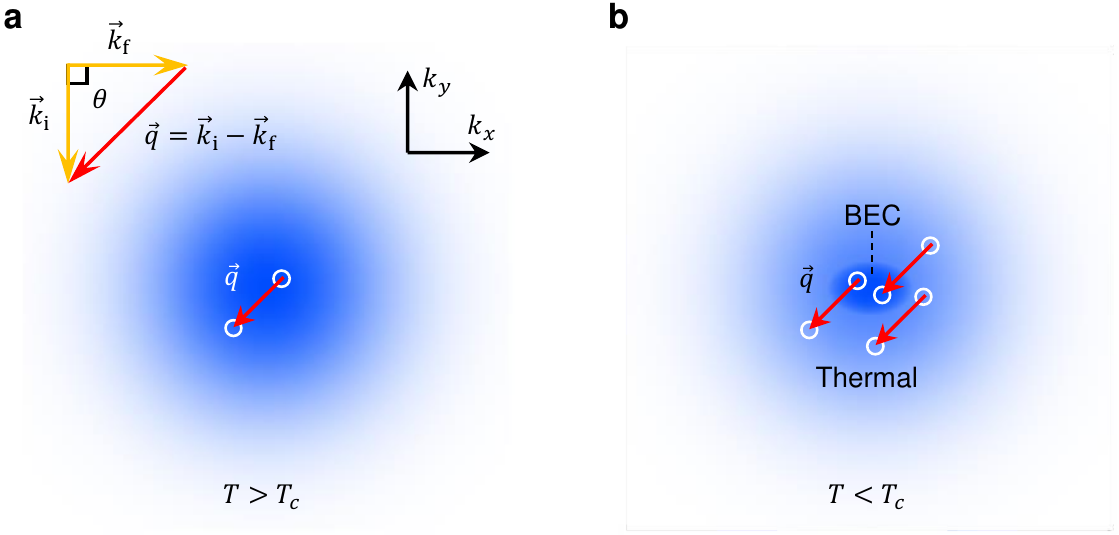}
\caption{\label{fig1} \textbf{Schematic for bosonic stimulation in light scattering.} Shown are momentum space pictures of the cloud above and below the phase transition. When a photon is scattered, the photon recoil changes the momentum of the atom by $\vec{q}$ (red arrow). (a) Above the phase transition ($T>T_c$), both the initial and final momentum states are in the thermal cloud. (b) Below the phase transition ($T<T_c$), a condensate emerges around zero momentum, and scattering between the condensate and the thermal cloud is also possible. In either case, light scattering is enhanced if the final state is occupied.}
\end{figure}

\begin{figure*}
\includegraphics[width=2\columnwidth]{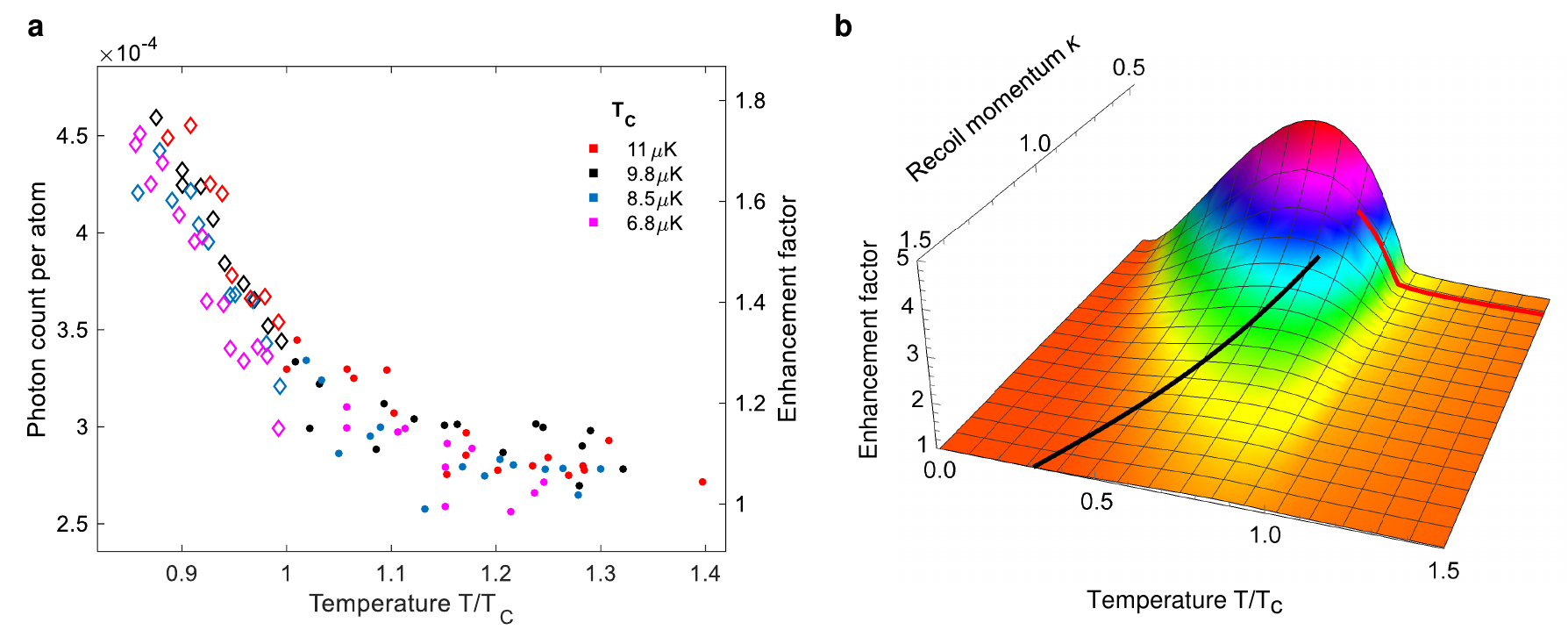}
\caption{\label{fig2} \textbf{Bosonic stimulation of light scattering in a trapped $^{23}$Na gas across the phase transition.} (a) Experimental data of light scattering versus temperature for different (normalized) recoil momentum $\kappa=\sqrt{E_\text{rec}/k_BT_c}$. The dots (diamonds) represents the data above (below) the phase transition. Data points here are each averaged over 12 to 16 samples, and the uncertainty is reflected by the scattering of the data. (b) Theoretical prediction of the bosonic enhancement factor for a harmonically trapped non-interacting gas. The red trajectory corresponds to the traces in (a), while the black trajectory corresponds to the trace when an almost pure condensate is prepared at low density and compressed.}
\end{figure*}

The principle behind bosonic stimulation in light scattering is illustrated in Fig.~\ref{fig1}. Light scattering is a two-photon process where an atom absorbs a photon (with momentum $\vec{k}_\text{i}$) from the probe beam and emits a scattered photon (momentum $\vec{k}_\text{f}$). The momentum of the atom changes due to the photon recoil by $\vec{q}=\vec{k}_\text{i}-\vec{k}_\text{f}$ (Fig.~\ref{fig1}a). 
In general, the (normalized) light scattering rate is given by the structure factor which can be obtained in the local density approximation by averaging over the phase-space~\cite{shuve2009enhanced}:
\begin{equation}
\label{eq1}
    S(\vec{q})= 1\pm \langle n(\vec{r},\vec{p}+\vec{q})\rangle,
\end{equation}
with the plus (minus) sign corresponding to bosons (fermions). The phase-space average is weighted by the occupation number $n(\vec{r},\vec{p})$. In the limit of zero recoil momentum, the second term reduces to the average phase-space density (PSD). This term becomes $1$ for fermions at zero temperature, predicting complete Pauli blocking~\cite{doi:10.1126/science.abi6153}. For bosons, the physics is richer because of the presence of the BEC phase transition. For a non-degenerate thermal gas, the averaged PSD is $n\lambda_{t}^3/2^{3/2}$ ($n$ is the density and $\lambda_{t}$ the thermal de Broglie wavelength). Extrapolating this formula to the BEC phase transition with the criterion $n\lambda_{t}^3=\zeta(3/2) \approx 2.612$ gives an enhancement factor of $1.92$ (where $\zeta$ is the Riemann zeta function). However, we will show that the enhancement factor at the transition temperature can diverge (in free space) or is limited to 1.37 for harmonic confinement.

The phase-space occupation number $n(\vec{r},\vec{p})$ in Eq.~\ref{eq1} is given by the Bose-Einstein occupation number in the thermal cloud
$n_\text{th}=\left[\exp\left[\left(\epsilon_{\vec{p}}+U(\vec{r})-\mu\right)/k_BT\right]-1\right]^{-1}$. The presence of a non-interacting condensate adds a $\delta$-function in phase space: 
\begin{equation}
\label{eq2}
    n(\vec{r},\vec{p})=
    \begin{cases}
     n_\text{th}(\vec{r},\vec{p}) & \text{$T\geq T_c$}\\
      n_\text{th}(\vec{r},\vec{p})|_{\mu=0}+h^3N_0\delta(\vec{r},\vec{p}) & \text{$T<T_c$}
    \end{cases}.       
\end{equation}
Here $\epsilon_{\vec{p}}=p^2/2m$, $U(\vec{r})$ and $\mu$ represent the kinetic energy, potential energy, and chemical potential, respectively. $N_0=f(T/T_c)N$ is the number of atoms in the condensate with $f$ being the condensate fraction.
 
\begin{figure*}
\includegraphics[width=2\columnwidth]{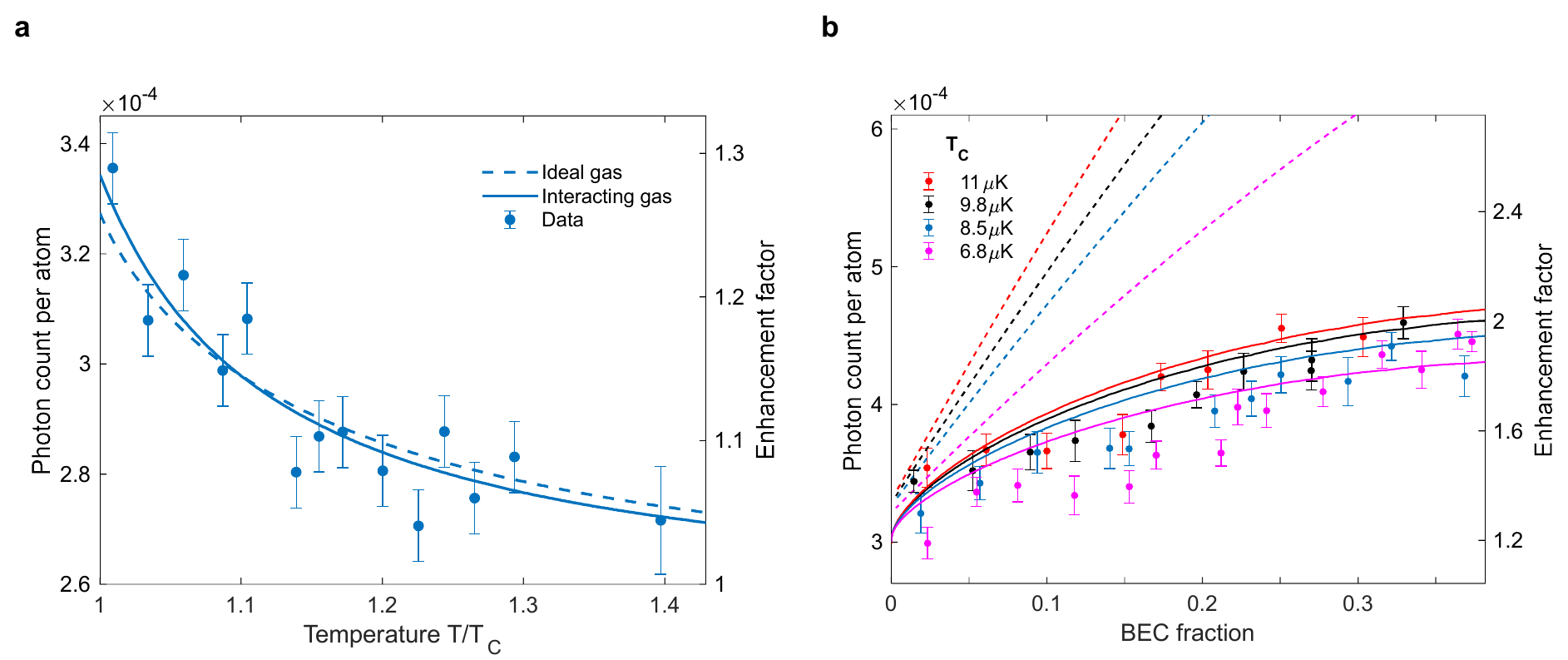}
\caption{\label{fig3} \textbf{Bosonic enhancement of light scattering above and below the BEC phase transition.} (a) Above the phase transition, the scattering happens within the thermal cloud. The dashed curve represents the theoretical prediction for ideal gas while the solid curve represents the interacting theory. The data is identical to the data in Fig.~\ref{fig2} with $T>T_c$ but averaged over nearby points. The error bars throughout the whole paper are purely statistical and reflect one standard error of the mean. 
(b) Below the phase transition, the scattering between BEC and thermal cloud dominates. 
The dashed curves are predictions for ideal gas, while the solid curves used semi-ideal gas approximation and the modified pair correlation function. 
The data is identical to the data in Fig.~\ref{fig2} with $T<T_c$. }
\end{figure*}

Above $T_c$, scattering occurs between two thermally occupied states (Fig.~\ref{fig1}a). The scattering rate is enhanced as the temperature approaches $T_c$ because of the increase of $n_\text{th}(\vec{r},\vec{p})$ at low momenta. Below the transition temperature, scattering into and out of the condensate becomes possible (Fig.~\ref{fig1}b). 
Combining the expression for $n(\vec{r},\vec{p})$ with Eq.~\ref{eq1}, the total scattering rate below $T_c$ reads: 
\begin{equation}
\label{eq3}
    S(\vec{q})=1+2fn_\text{th}(0,\vec{q})+\langle n_\text{th}(\vec{r},\vec{p}+\vec{q})|_{\mu=0}\rangle.
\end{equation}
The first term represents the contribution of single-particle Rayleigh scattering events, the second term is the condensate-to-thermal scattering events, while the third term describes thermal-to-thermal scattering events. 
Already for moderate condensate fractions, the scattering is dominated by scattering events involving the condensate. 
As the temperature becomes lower, the condensate fraction $f$ increases but the occupation number $n_\text{th}(0,\vec{q})$ ($n_\text{th}(0,\vec{q})\approx k_BT/ E_\text{rec}$ for $k_BT\gg E_\text{rec}$) in the thermal cloud decreases. As a result, the total scattering rate first increase and then decrease as the temperature is further decreased below $T_c$. Ultimately at zero temperature, all atoms are in the condensate and bosonic stimulation is absent because of zero occupation in the final states. Fig.~\ref{fig2}b shows the theoretical calculation for a non-interacting Bose gas in a harmonic trap. 

Bosonic enhancement monotonically increases when the (normalized) recoil momentum $\kappa=\sqrt{ E_\text{rec}/k_BT_c}$ decreases (where $E_\text{rec} = \hbar^2k^2/m$ is the atomic recoil energy). In a harmonic trap, for $\kappa \rightarrow0$, the enhancement above the BEC phase transition is bounded by $\zeta(2)/\zeta(3)\approx 1.37$, while the enhancement factor below $T_c$ diverges as $1/\kappa^2$. In contrast, in free space (3D box potential) the enhancement factor diverges already above $T_c$ as $\pi^{3/2}\kappa^{-1}/\zeta(3/2)$. This major difference illustrates that in a harmonic trap at $T_c$, only a small volume is at or near the conditions for the phase transition. Derivation of these results, as well as a generalization to arbitrary power-law potentials, can be found in the Supplementary Materials~\cite{SM}. 

\begin{figure*}
\includegraphics[width=2\columnwidth]{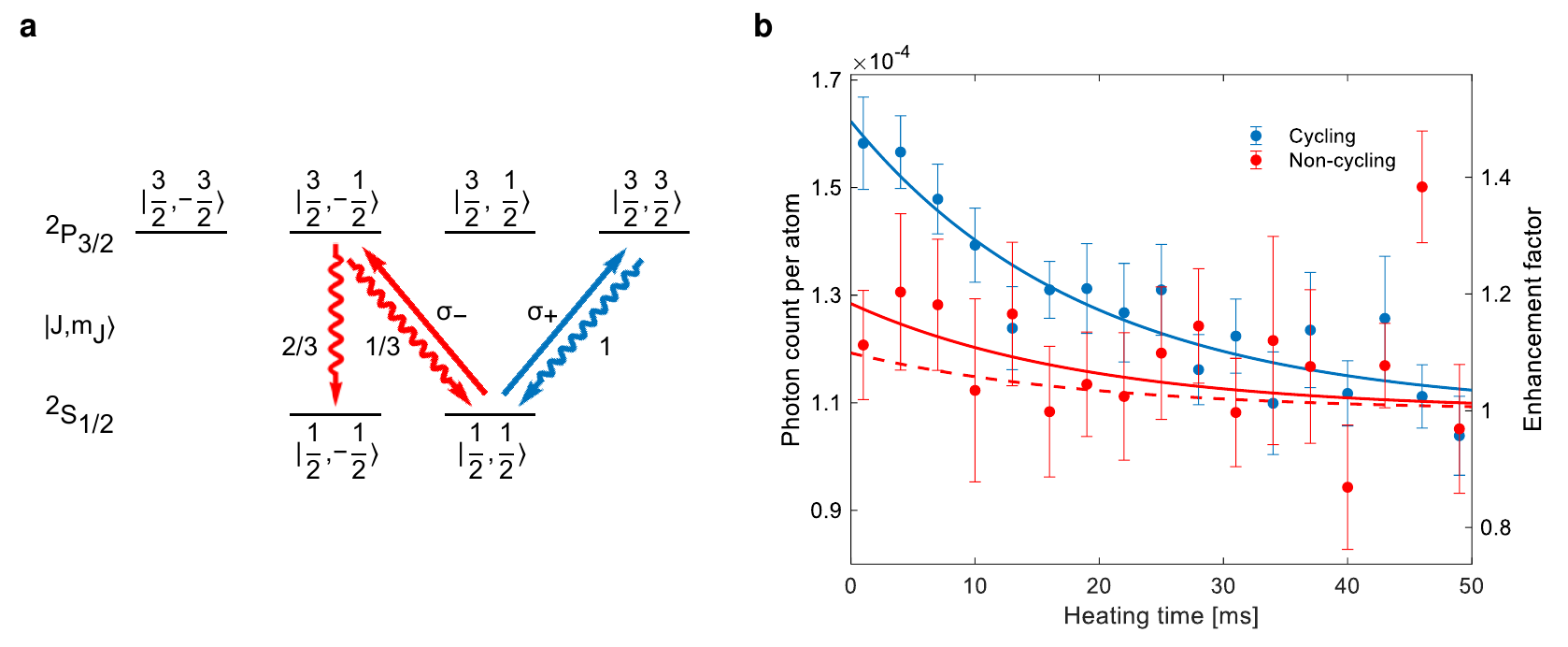}
\caption{\label{fig4} \textbf{Bosonic enhancement for Rayleigh and Raman scattering.} (a) For large detuning, the atom's internal states are labelled by $\ket{J,m_J}$ where $J$ represents the total angular momentum of the electron and $m_J$ its projection along the quantization axis. The blue (red) straight arrow represents the excitation paths for probe light with $\sigma_+$ ($\sigma_-$) polarization, while the wavy arrows represent the emitted photons with branching ratio labeled on the side. (b) Observed photon scattering for $\sigma_+$ (blue) and $\sigma_-$ (red, scaled by $9/5$) polarized probe light. The solid blue curve is a fit to the $\sigma_+$ data (using an exponential function as a convenient trial function), the red dashed light is obtained from the blue fit and Eq.~\ref{eq4}. The solid red line is a fit to Eq.~\ref{eq4} with adjustable $\gamma$. Each data point is the average over 12 samples.}
\end{figure*}

In the experiment, around $4\times 10^5$ partially condensed $^{23}$Na atoms in hyperfine state $\ket{F=2, m_F=2}$ are prepared in an optical dipole trap~\cite{SM}. The cloud is then heated up to different temperatures by modulating the trap intensity while keeping the atom number almost constant. Light is scattered by exciting the cycling transition with $\sigma_+$ polarized light. The optical setup is described in our previous work~\cite{doi:10.1126/science.abi6153}. For the measurements to be perturbative, the scattering rate was kept below 0.14 photon per atom during 4 ms. 
We observed light scattering for $4$ different densities over a range of temperatures (Fig.~\ref{fig2}a). The density is controlled by adjusting the depth of the optical dipole trap and determines $T_c$ and $\kappa$. The red line in Fig.~\ref{fig2} shows the parameters of these measurements (the traces for the different values of $T_c$ overlap). 

The data in Fig.~\ref{fig2}, even without quantitative analysis, show the salient features of light scattering for bosons: above the phase transition, there is a modest enhancement in scattering due to the increase in phase space density, as the temperature is reduced. Once the condensate forms, the enhancement is more substantial due to scattering into and out of the condensate. We also observed the absence of enhanced light scattering for an almost pure Bose-Einstein condensate. However, due to the strong increase of three-body loss at high density, we could prepare such condensates only at lower densities in more weakly confining traps where the maximum enhancement is small due to the larger value of $\kappa$. When the cloud was compressed to
higher densities (shown as the black trajectory in Fig.~\ref{fig2}b), an increase of light scattering was observed, qualitatively consistent with the prediction~\cite{SM}. 



For temperatures above $T_c$, light scattering occurs within the thermal distribution. In Fig.~\ref{fig3}a, we compare the results with the predictions for an ideal gas and an interacting gas, without any free parameters except the overall scaling. Data with different $T_c$ are binned together because of their small difference in $\kappa$. 
Usually, thermal Bose gases away from Feshbach resonances are treated as non-interacting. However, at our high densities, the mean-field repulsion is relevant even for thermal clouds; it softens the trapping potential and increases the bosonic enhancement by up to $50\%$ because of the higher density of states at low energy~\cite{SM}. Bosonic stimulation is reduced by a second effect of interactions: the reduction of the pair correlation function suppresses the enhancement by a factor $1-8\sqrt{2}a/\lambda_t\approx 0.6$ at high temperatures~\cite{PhysRevA.59.4595, SM}. Close to the phase transition, the thermal de Broglie wavelength $\lambda_t$ is effectively replaced by the correlation length $\xi$. The divergence of $\xi$ at the phase transition will diminish the suppression caused by pair correlation. To the best of our knowledge, the suppression of bosonic stimulation by repulsive interactions has not been discussed before. In the supplement, we discuss some evidence for the effect of pair correlations and its reduction near criticality. Fig.~\ref{fig3}a demonstrates that the two effects of interactions almost cancel each other.

For temperatures below the BEC phase transition, the scattering between the condensate and the thermal cloud dominates the bosonic enhancement. For non-interacting Bose gas, the enhancement factor increases almost linearly with condensate fraction close to the critical point: $ S(\vec{q})\approx1+2fk_BT/ E_\text{rec}$ (dashed curves in Fig.~\ref{fig3}b). 
However, because of the mean-field repulsion between the thermal cloud and the condensate, the density of thermal atoms inside the condensate volume is strongly reduced~\cite{PhysRevA.58.2423} diminishing the bosonic enhancement. The enhancement $S(\vec{q})-1$ is thus reduced by the factor $(\int d\vec{r} n_0(\vec{r})n_\text{th}^\text{(int)}(\vec{r},\vec{q}))/(N_0n_\text{th}(0,\vec{q}))$, with $n_0(\vec{r})$ being the density distribution of the interacting BEC, and $n_\text{th}^\text{(int)}(\vec{r},\vec{q})$ the phase-space occupation number for the interacting thermal cloud. The experimentally observed enhancement is reduced by up to a factor of 3 compared to the non-interacting theory (Fig.~\ref{fig3}b). We model the effect of interactions using the semi-ideal gas approximation~\cite{PhysRevA.58.2423} which assumes an interacting condensate and repulsion of the thermal cloud by the condensate, and include the modified pair correlation function~\cite{PhysRevA.59.4595} (Fig.~\ref{fig3}b). Note that the theory curves in Fig.~\ref{fig3}b has no free parameters (the overall scaling was the same as Fig.~\ref{fig3}a). Our model explains most of the reduction in light scattering compared to the predictions for a non-interacting gas. The remaining difference could be due to the backaction from the thermal cloud to the BEC which will further reduce their overlap which can be approximately treated using a Hartree-Fock approximation~\cite{SM}. However, a fully consistent treatment of the interacting bose gas is challenging both theoretically and numerically.

Until now, we have assumed a two-level system, realized by exciting the initial $\ket{F=2, m_F=2}$ hyperfine state on a cycling transition with $\sigma_+$ polarized light (see Fig.~\ref{fig4}a, blue). 
We now discuss the situation when the atoms are excited with light of different polarization. In the limit of detunings much larger than the hyperfine interaction in the excited state, the nuclear hyperfine structure can be neglected since the nuclear spin state does not change during the light scattering. Therefore, we realize the simplified level diagram shown in Fig.~\ref{fig4}a.


When the atoms are initially prepared in a single internal state, only Rayleigh scattering shows bosonic enhancement. Raman scattering (or optical pumping) populates initially empty states and is not enhanced. We therefore expect the enhancement factor $\eta$ to be reduced to $\gamma \eta +(1-\gamma)$ where $\gamma$ represents the branching ratio (ratio of Rayleigh scattering to Raman scattering). To describe our experimental situation, we express the observed scattering rate for $\sigma_+$ light as $R_+=\eta R$, with $R$ representing the Rayleigh scattering rate without any enhancement. The expected observed scattering rate for $\sigma_-$ light is then predicted as
\begin{equation}
\label{eq4}
    R_-=\frac{1}{3}(\gamma\eta R+2(1-\gamma)R).
\end{equation}
Here the prefactor $1/3$ comes from the reduced matrix element, the branching ratio $\gamma=1/3$, and the extra factor of two comes from the different angular distributions for the scattered $\sigma$ and $\pi$ light. 
By comparing Eq.~\ref{eq4} in the high temperature and low-temperature limit, we find the enhancement factor for $\sigma_-$ polarized light to be $(4+\eta)/5$. 
The results presented in Fig.~\ref{fig4}b confirm that the bosonic enhancement for $\sigma_-$ probe light is always below the enhancement for $\sigma_+$. We can use Eq.~\ref{eq4} and regard $\gamma$ as a fitting parameter which characterizes the bosonic enhancement. 
The best fit to the data gives $\gamma= 0.5(2)$, and is consistent with the theoretical prediction ($\gamma= 1/3$) within the experimental uncertainty (which is larger for $\sigma_-$ scattering due to the small number of photons observed). If the light scattering is non-perturbative, the population can build up in the initially empty state via optical pumping which will then lead to bosonic enhancement of the Raman transition which has been observed in the form of Raman superradiance~\cite{PhysRevA.69.041601,PhysRevA.69.041603}. 


In conclusion, we have shown how quantum statistics and interaction modifies the optical properties of a Bose gas. We have worked in a regime where the effect of quantum statistics dominate over other effects such as dipole-dipole interactions~\cite{PhysRevA.94.023612}, superradiance~\cite{inouye1999superradiant}, light-assisted collisions~\cite{RevModPhys.71.1}, multiple scattering and atomic lensing. 
Besides the fundamental interest, understanding light scattering is crucial for quantitative diagnostics of bosonic systems. 
For future work, it is promising to study bosonic enhancement of light scattering in a box potential~\cite{box} because of the absence of density inhomogeneity. The large extension of the critical region will give much stronger enhancement and enable studies of the $1/\kappa$ divergence of the enhancement factor. In addition, in a box potential, interactions can affect light scattering only through the modification of the pair correlation function. It should be possible to clearly observe how bosonic enhancement is reduced above the phase transition, but not right at the critical point. 
 Furthermore, for strong repulsive interactions near a Feshbach resonance, light scattering in bosons could possibly be even suppressed below the single-particle Rayleigh rate, in analogy with Pauli blocking for fermions. Both Pauli blocking and strong interactions reduce the pair correlation function. More generally, light scattering provides a way to measure the static structure factor which could be used to characterize strongly interacting systems, including systems with strong dipolar interactions which are anisotropic and long-range.

\vbox{}
\vbox{}
\noindent \textbf{\large Acknowledgments} \\
We thank Julius de Hond and Pierre Barral for comments on the manuscript. \textbf{Funding:} This work is supported by NSF through the Center for Ultracold Atoms and Grant No. 1506369, and from a Vannevar-Bush Faculty Fellowship. \textbf{Author Contributions:} 
All authors contributed to the concepts of the experiment, Y.-K.L. and Y.M performed the experiment, Y.-K.L. analyzed the data and performed the theoretical calculations, Y.-K.L., Y.M and W.K. wrote the paper. 
\textbf{Competing interests:} The authors declare no competing interests. 
\textbf{Data Availability:} The data that support the plots within this paper and other findings of this study are available from the corresponding authors upon reasonable request.

\vbox{}
\noindent\textbf{\large Supplementary Materials}\\
Materials and Methods\\
Supplementary Text\\
Figs. S1-S6\\
Table S1-S2\\
References (S1-S9)

\clearpage
\renewcommand{\citenumfont}[1]{S#1}
\renewcommand{\bibnumfmt}[1]{[S#1]}
\renewcommand{\theequation}{S\arabic{equation}}
\renewcommand{\thesection}{S\arabic{section}}
\renewcommand{\thefigure}{S\arabic{figure}}
\renewcommand{\thetable}{S\arabic{table}}
\setcounter{figure}{0}
\setcounter{table}{0}

\widetext
\begin{center}
\textbf{\large Supplemental Materials: Observation of bosonic stimulation in light scattering}
\end{center}

\section{Experimental details}
\textbf{Sample preparation}
Around 1 billion sodium atoms are loaded into a magneto-optical trap (MOT) and transferred into a quadrupole magnetic trap with an optical plug where they are cooled by microwave evaporative cooling. Subsequently, around $10^7$ atoms in hyperfine state $\ket{F=2,m_F=2}$ are transferred into a single-beam 1064 nm optical dipole trap. The beam waist and intensity of the trapping beam are controlled by a variable-size iris and an acousto-optic modulator, respectively. The beam waist was initially chosen to be big to increase the loading efficiency into a large trap volume. Because the bosonic enhancement effect requires $k_BT_c\gg E_\text{rec}$, the atomic gas is compressed to a higher density with a higher $T_c$ by fully opening up the iris. Decreasing the intensity of the optical trapping beam induces evaporative cooling to a Bose-Einstein condensed cloud with around $4\times10^5$ atoms and around 30\% condensate fraction.
The final $T_c$ can be adjusted from $6.8\mu K$ to $11\mu K$ by changing the final power of the optical trap. Subsequent heating of the cloud was done by sinusoidal modulating the power of the optical trapping beam by 50\% at 1 kHz frequency with variable duration (from 1 ms to 49 ms). 

\textbf{Light scattering and collection}
The probe light was generated from an independent 1178 nm laser with frequency doubling. The frequency of the probe light was red detuned from the D2 transition of sodium by $\Delta=25.7 \text{ GHz}$. The large detuning was chosen to avoid multiple scattering of photons, to minimize the lensing effect and photo-association loss. In addition, since the detuning is much larger than the hyperfine interaction in the electronically excited state ($\Delta\gg A_\text{HFS}=18.5\text{ MHz}$), the nuclear spin of the atom is preserved during the light scattering. The $1/e^2$ diameter of the probe beam is $3.3$ mm in the atoms' plane. The intensity of the beam was stabilized to avoid shot-to-shot fluctuations.

Fluorescence light is collected at a 90-degree angle relative to the probe beam with a collection efficiency of around 0.3\%. To detect the small number of scattered photons, we used an EMCCD camera with quantum efficiency of 97\% and readout noise of 3 - 5$e^-$ per pixel. Hardware binning of pixels was implemented to make sure the photon shot noise dominates over the readout noise. Background noise was suppressed by choosing a region of interest only slightly larger than the image of the cloud. 



For the light scattering in Figs.~\ref{fig2} and~\ref{fig3} at most 0.14 photons were scattered per atom. The pulse duration was 4 ms. For the data in Fig.~\ref{fig4}, since the $\sigma_-$ polarized probe light will cause optical pumping, we used weaker pulses with at most 0.05 (0.02) photon per atom for Rayleigh (Raman) scattering and a pulse duration of 3 ms.

\begin{figure}
\includegraphics[width=0.5\columnwidth]{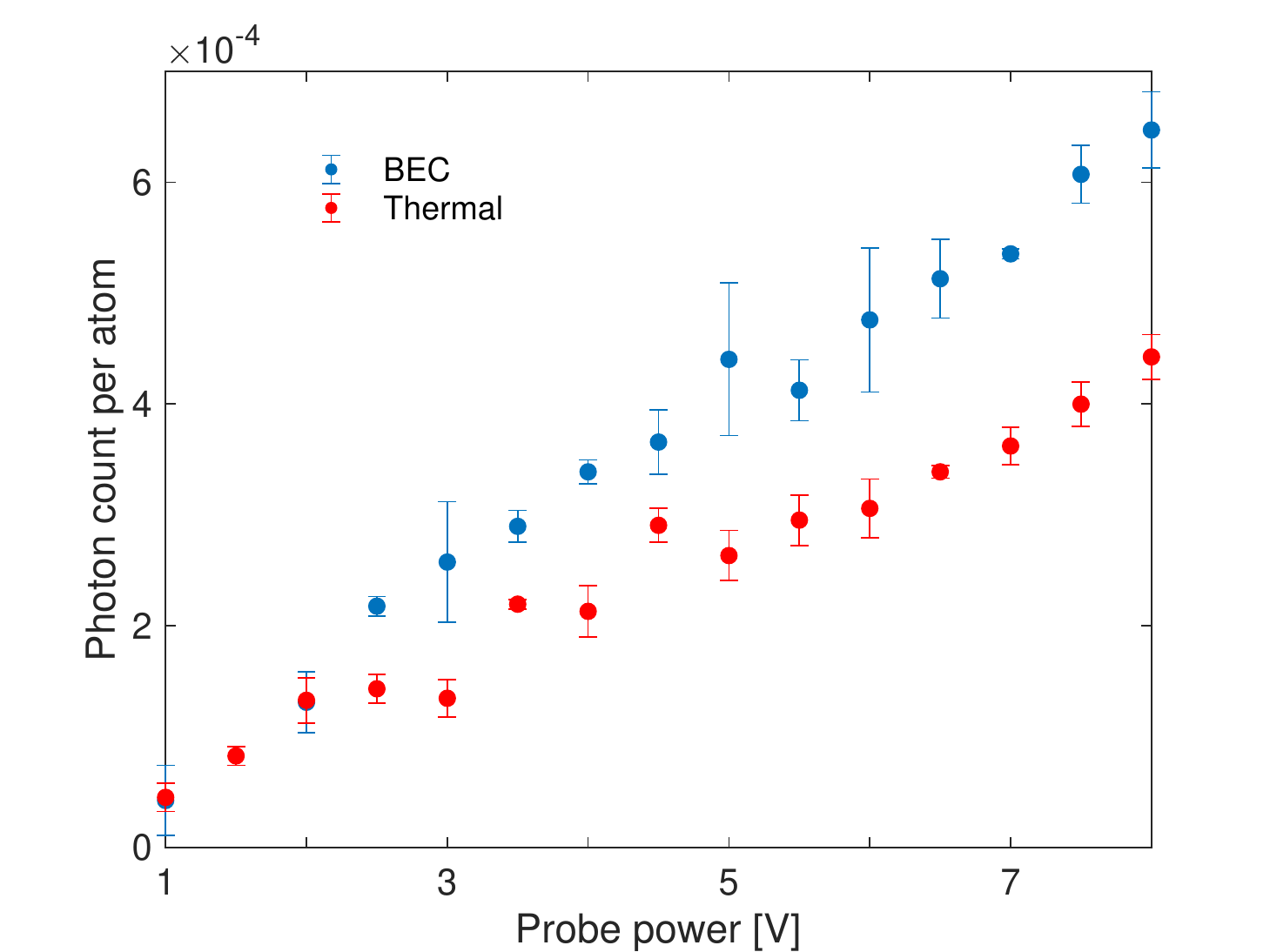}
\caption{\label{linear} \textbf{Linearity check for light scattering.} Scattered photons are measured for different probe beam powers. The blue (red) data points represents a cloud with (without) BEC. The absence of nonlinearity shows that we are working in the perturbative regime. Data points here are each averaged over 3 samples. }
\end{figure}

We verified that the atoms were probed in the perturbative regime where optical pumping and nonlinear effects such as superradiance are absent, by measuring the scattered light as a function of probe beam power (Fig.~\ref{linear}). 
The measurement was done for both a cloud with and without BEC. The linear behavior shows that the light scattering is indeed perturbative.

\textbf{Thermometry}
The temperature of the cloud was determined from time-of-flight absorption measurement. The fitting of absorption images was done using the Bose function for the thermal cloud and Thomas-Fermi distribution for the BEC~\cite{ketterle1999making}. Since the atoms were released from the trap 8 ms after the light pulse, we had to account for a possible change in temperature during this hold time. The hold time was chosen to allow the atoms to thermalize, and was also necessary for a mechanical camera shutter to open which was necessary to block the probe beam during the light scattering.

Because of the high density of the cloud, the lifetime of the cloud due to three-body recombination was around 15 ms. Heating due to particle loss (``anti-evaporation'') during the light scattering and the extra hold time was therefore non-negligible. We accounted for this effect by measuring the temperature for identical cycles, but without the light scattering pulse and the extra hold time. This temperature (averaged over multiple cycles) was measured for each sample heating time and used for the temperature axes in Figs.~\ref{fig2} and~\ref{fig3}. 

\textbf{Properties of the high-density cloud}
Significant Bose enhancement of light scattering at large angles requires the critical temperature to be higher than the recoil energy. Therefore, the density $n$ of our sample needed to be around $10^{15} \text{ cm}^{-3}$. Since the three-body loss rate coefficient for thermal sodium atom in $F=2$ states at zero magnetic field is $9.6\times10^{-29}\text{ cm}^6\text{/s}$, the lifetime of the cloud is limited to around 15 ms. Another consequence of the high density is the short mean free path ($l=1/n\sigma\approx1\mu m$ where $\sigma$ is the cross-section for elastic collisions). Since the typical size of the cloud is around $1\mu m$ in the radial direction and $30\mu m$ in the axial direction, this implies slow hydrodynamic transport in the axial direction. The typical trap frequencies in the radial (axial) direction are 9.5 kHz (290 Hz). Due to hydrodynamic collisional effects, the equilibrium time scale in the axial direction is much slower than the trap period and can potentially lead to non-fully equilibrated clouds in combination with fast three-body loss. More precise measurement of bosonic light scattering should therefore be done at lower densities and smaller scattering angles, or by choosing atoms with even lower three-body loss rate (e.g., $^{23}\text{Na}$ in the $F=1$ state or $^{164}\text{Dy}$). 

\section{Light scattering for a pure BEC and adiabatic compression} 
\begin{figure}
\includegraphics[width=0.5\columnwidth]{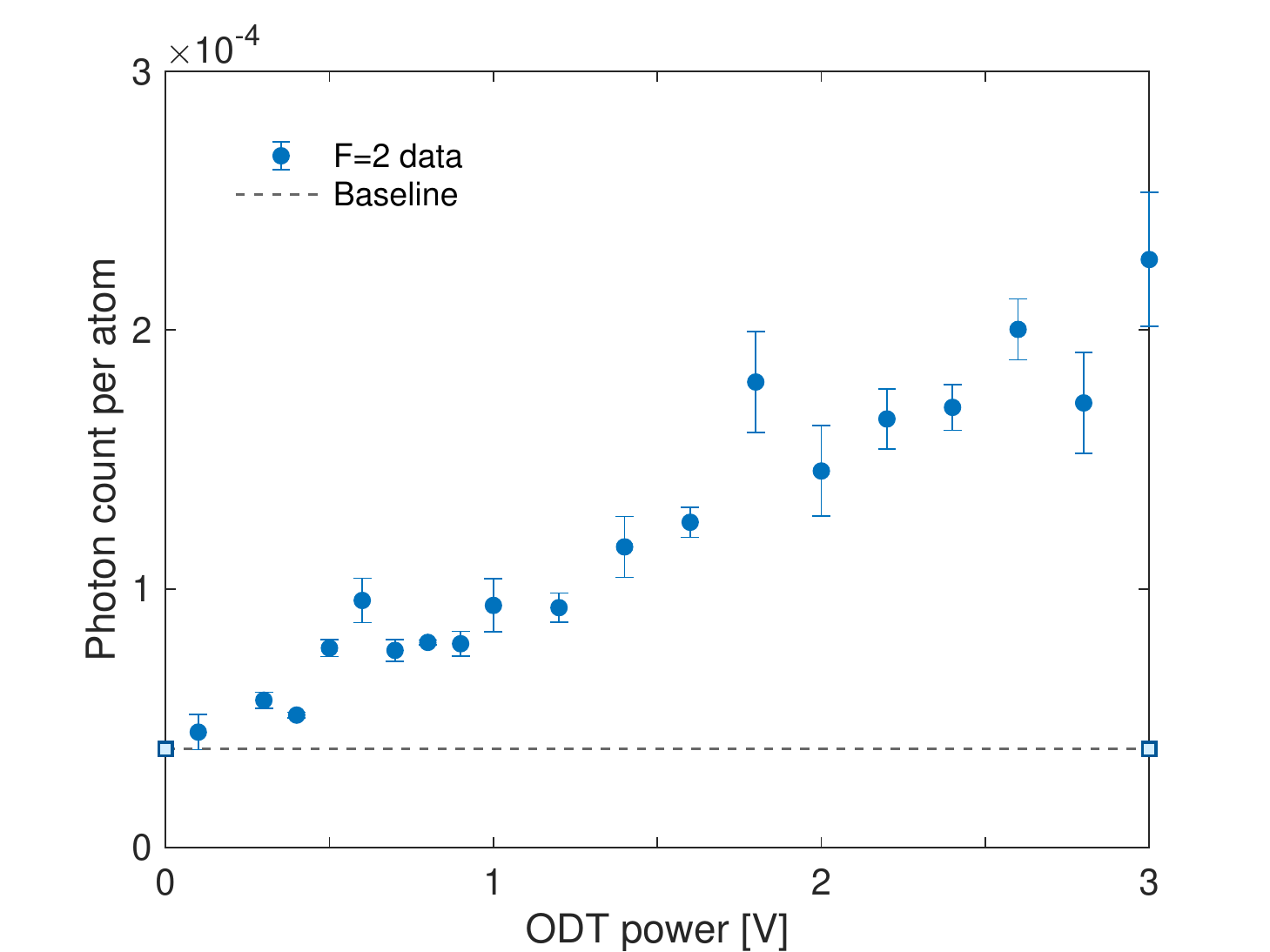}
\caption{\label{compress} \textbf{Light scattering after compressing the cloud.} The photon scattering signal was measured for different final trap depths. The dashed line represents the signal observed for a non-degenerate cloud with no bosonic enhancement. Each data point was averaged over 6 samples. }
\end{figure}

A pure Bose-Einstein condensate shows no bosonic enhancement of light scattering. We therefore studied light scattering using an almost pure condensate, and then compressed the condensate adiabatically by ramping up the power of the optical trap. 
An almost pure BEC could be prepared only in a shallow trap.
 Subsequently, the trap power was ramped up to increase the density and therefore $T_c$. Light scattering was measured for different final trap depths. Without compression, the photon count rate is identical to the signal for a non-degenerate thermal gas (Fig.~\ref{compress}), demonstrating the absence of bosonic enhancement in a pure condensate. However, for the low values of $\kappa=1.5$, the maximum predicted enhancement for a mixed cloud is only 9\%. For higher densities, bosonic enhancement of light scattering was observed, in qualitative agreement with the theory. During compression, the cloud approximately followed the black trajectory in Fig.~\ref{fig2}. Due to losses and heating during compression, the condensate fraction decreased, and light scattering was enhanced. Due to the heating, we were not able to map out the region of low $T/T_c$ and low $\kappa$, where the enhancement would increase with temperature.

 
\section{Theory for the bosonic enhancement factor}
The bosonic enhancement factor approaches zero in the limit of low density or large recoil, $\kappa \rightarrow \infty$, and is maximized in the high-density limit $\kappa=0$ where the normalized light scattering is given by the $S(\vec{q}=0)$ at zero momentum. We now rigorously calculate $S(0)$ for power-law densities of state $g(\epsilon)=A\epsilon^x$. For a generic power-law trapping potential $V(r)\propto r^\alpha$ in $d$ dimensional space, the exponent $x$ in the density of state reads~\cite{PhysRevA.59.3109}:
\begin{equation}
    x=\frac{d}{2}+\frac{d}{\alpha}-1.
\end{equation}
The box potential has $\alpha=\infty$, and in three dimensions $x=1/2$.

Rewriting the integrals in Eq.~\ref{eq1} in terms of energy $\epsilon$, we have:
\begin{equation}
    S(\vec{q}=0)=1+\frac{\int_0^\infty d\epsilon n(\epsilon)^2g(\epsilon)}{\int_0^\infty d\epsilon n(\epsilon) g(\epsilon)}.
\end{equation}
Here $n(\epsilon)=(z^{-1}\exp(\beta\epsilon)-1)^{-1}$ is the occupation number with $z=\exp(\beta\mu)$ being the fugacity and $\beta=1/k_BT$. 

The integrals have analytic solutions:
\begin{equation}
    \int_0^\infty d\epsilon n(\epsilon) g(\epsilon)=A\Gamma(1+x)g_{1+x}(z)/\beta^{1+x},
\end{equation}
\begin{equation}
    \int_0^\infty d\epsilon n(\epsilon)^2 g(\epsilon)=A\Gamma(1+x)(-g_{1+x}(z)+g_x(z))/\beta^{1+x}.
\end{equation}
Here $g_n(z)$ represents the $n$th order polylogarithm function. 
Finally, we get a simple expression for the bosonic enhancement factor:
\begin{equation}
     S(0)= \frac{g_x(z)}{g_{1+x}(z)}.
\end{equation}

It can be shown that ${g_x(z)}/{g_{1+x}(z)}$ has its maximum at the BEC phase transition $z=1$. The behavior of the polylogarithm functions at $z=1$ divides the physics into three regimes depending on the value of $x$. The results are summarized in table~\ref{tb1}.

When $x\leq0$, both $g_x(z)$ and $g_{1+x}(z)$ diverge at $z=1$. The divergence of $g_{1+x}(z)$ implies the absence of the BEC phase transition. Therefore, $z<1$ for all temperatures, and the enhancement factor $g_x(z)/g_{1+x}(z)$ is always finite.

When $0<x\leq1$, $g_x(z)$ diverges at $z=1$ but $g_{1+x}(z)$ remains finite. This implies the presence of a BEC phase transition and (for $\kappa=0$) an unbounded light scattering rate with a divergent enhancement factor at the BEC phase transition point. 

When $x>1$, both $g_x(z)$ and $g_{1+x}(z)$ remains finite at $z=1$. This implies the presence of a BEC phase transition and an bounded light scattering rate. The maximum possible enhancement factor at the tranistion temperature is ${\zeta(x)}/{\zeta(1+x)}$.

\begin{table*}
\begin{tabular}{|c|c|c|c|}
\hline
Exponent $x$         & $x\leq0$       & $0<x\leq1$                    & $x>1$             \\ \hline
BEC phase transition & No             & Yes                           & Yes               \\ \hline
Bosonic enhancement & Bounded & Diverges as $\kappa^{2x-2}$ when $T\rightarrow T_c$ & Bounded by ${\zeta(x)}/{\zeta(1+x)}$ for $T\geq T_c$\\ \hline
\end{tabular}%
\caption{\label{tb1} \textbf{Bosonic enhancement factor for different exponent $x$ in the density of states.} In the limit of zero recoil momentum, the enhancement factor is bounded for $x\leq0$ and $x>1$ while diverging for $0<x\leq1$.}
\end{table*}

\begin{figure}
\includegraphics[width=0.5\columnwidth]{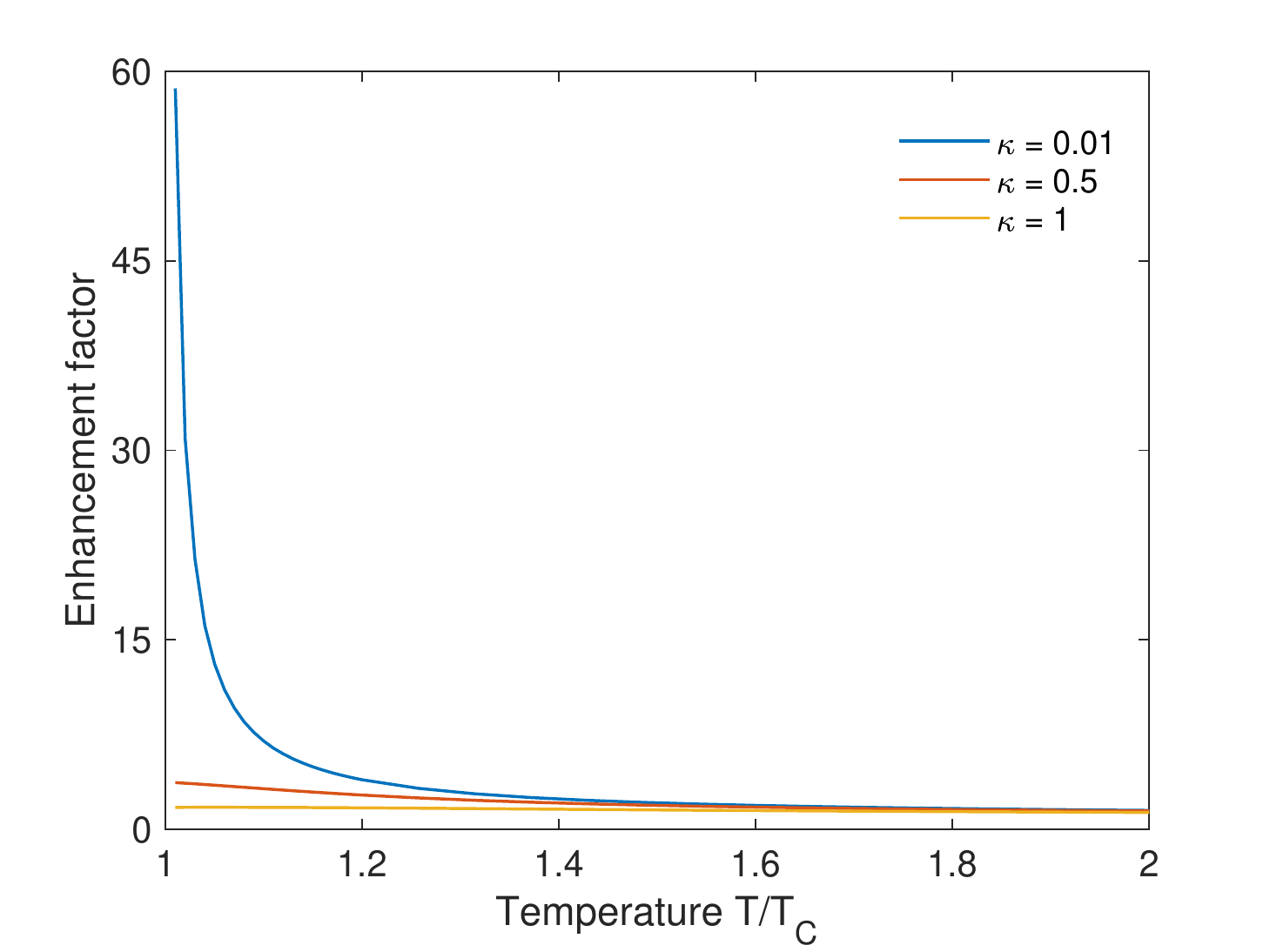}
\caption{\label{box} \textbf{Bosonic enhancement factor for an ideal gas in a 3D box potential for different recoil momenta $\kappa$.} At the phase transition point, the bosonic enhancement factor diverges as $1/\kappa$. }
\end{figure}

\begin{figure}
\includegraphics[width=0.5\columnwidth]{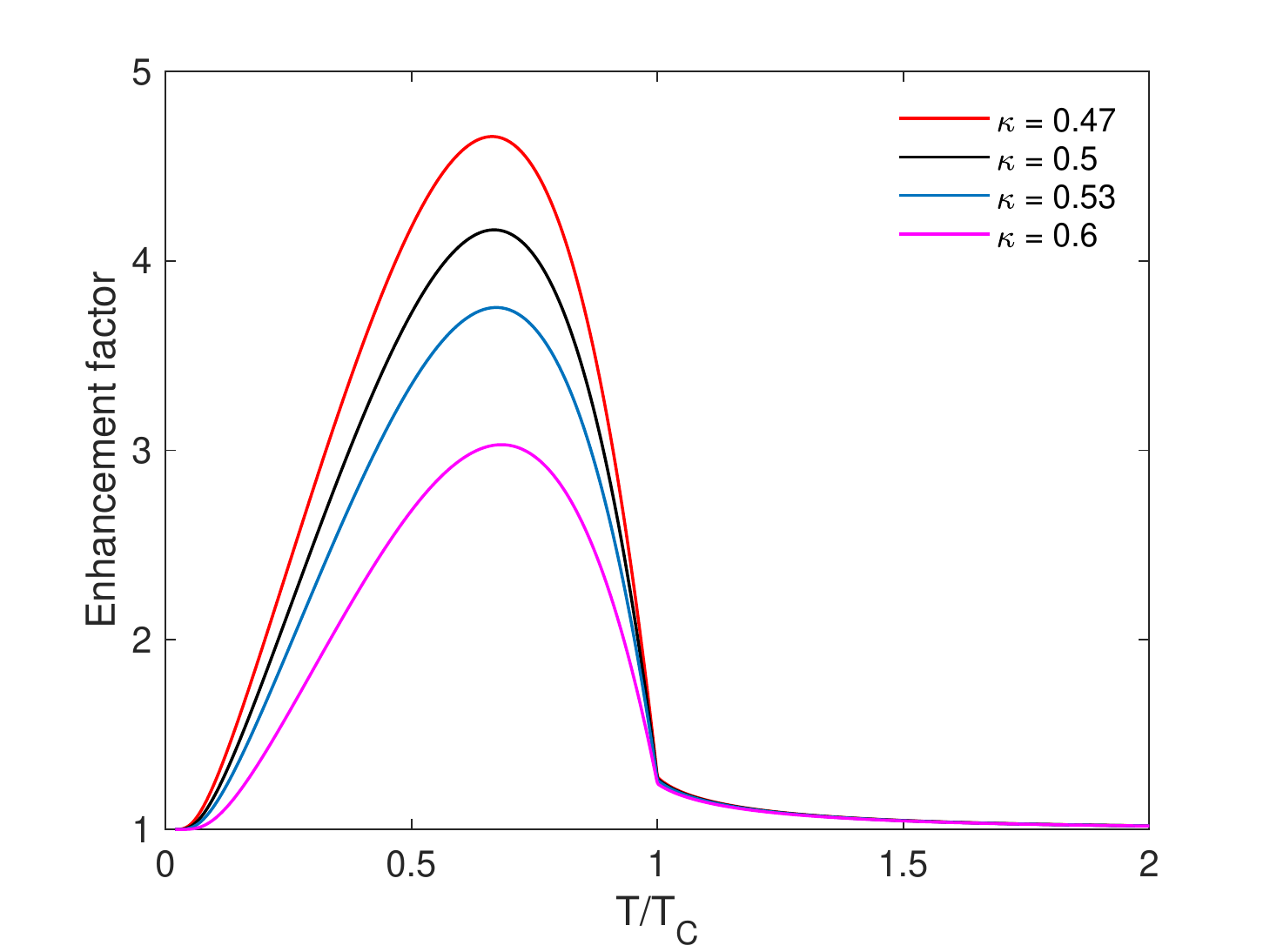}
\caption{\label{harmonic} \textbf{Bosonic enhancement factor for an ideal gas in a 3D harmonic trapping potential for different recoil momenta $\kappa$.} The enhancement factor is bounded at and above the phase transition, but will diverge as $1/\kappa^2$ below the phase transition.}
\end{figure}

For the relevant case of the 3D box potential ($x=1/2$), light scattering can diverge, whereas for the 3-dimensional harmonic trapping potential ($x=2$) the maximum enhancement factor is $\zeta(2)/\zeta(3)\approx 1.37$. Figs.~\ref{box} and~\ref{harmonic} illustrate those cases. 

To study the divergence of the enhancement factor near the phase transition for $0<x\leq1$, we now determine the behavior for small, but non-zero recoil momentum $\kappa$.

We start with the exact expression for the enhancement factor at finite recoil $\kappa$ at the phase transition:
\begin{equation*}
    S(q)=1+\frac{1}{\sqrt{\pi}\Gamma(x+1/2)\zeta(x+1)}\int_0^\infty da\int_{-\infty}^\infty dy\frac{1}{\exp(a+y^2)-1}\frac{a^{x-1/2}}{\exp(a+(y+\kappa)^2)-1}.
\end{equation*}
Here $y$ represents the dimensionless momentum in the recoil direction, and $a$ represents the total energy from all other dimensions in phase space. Since the enhancement factor diverges as $\kappa^{2-2x}$ for $0<x\leq1$, we need to calculate:
\begin{equation}
    I=\lim_{\kappa\rightarrow0}\int_0^\infty da\int_{-\infty}^\infty dy\frac{\kappa^{2-2x}}{\exp(a+y^2)-1}\frac{a^{x-1/2}}{\exp(a+(y+\kappa)^2)-1}.
\end{equation}

After transforming to polar coordinates $a=r^2\sin(\theta)^2$ and $y=-r\cos(\theta)$, one obtains $I=\lim_{\kappa\rightarrow0}\int_0^\infty dr\int_{0}^{\pi} d\theta f(r,\theta)$ with
\begin{equation}
  f(r,\theta)=\frac{\kappa^{2-2x}}{\exp(r^2)-1}\frac{2r(r\sin(\theta))^{2x}}{\exp(r^2-2r\kappa\cos(\theta)+\kappa^2)-1}.
\end{equation}
We first perform the integral over $r$. The function $f(r,\theta)$ has a branch cut from $r=0$ to $r=+ \infty$. It also has poles at $r_{\pm,n}=\kappa\cos(\theta)\pm\sqrt{-\kappa^2\sin(\theta)^2+2\pi ni}$ with $n$ being an integer. The residues at $r=r_{\pm.n}$ are:
\begin{equation}
  \text{Res}(f(r_{\pm,n})) = \pm\frac{2\kappa^{2-2x}\sin(\theta)^{2x}}{\exp((r_\pm)^2)-1}\frac{(r_\pm)^{2x+1}}{r_+-r_-}.
\end{equation}
In the limit of $\kappa\rightarrow0$, the residues vanishes if $n\neq0$. For $n=0$, the residues remain finite:
\begin{equation}
   \text{Res}(f(r_{\pm,0})) =  \pm\frac{2\kappa^{2-2x}\sin(\theta)^{2x}}{\exp(\kappa^2\exp(\pm2i\theta))-1}\frac{(\kappa\exp(\pm i\theta))^{2x+1}}{2i\kappa\sin(\theta)}\approx-\pm i(\exp(\pm i\theta)\sin(\theta))^{2x-1}.
\end{equation}
By using the Hankel contour in the complex $r$ plane, the integral $I$ can be related to the sum of all residues in the complex plane by:
\begin{equation}
I(1-\exp(2i\pi x))=2\pi i\sum_n\int_0^{\pi}d\theta\text{Res} (f(r_{\pm,n},\theta)),
\end{equation}
with the result:
\begin{equation}
I=\frac{2\pi i\int_0^{\pi} d\theta(- i(\exp(i\theta)\sin(\theta))^{2x-1}+ i(\exp(- i\theta)\sin(\theta))^{2x-1})}{1-\exp(2i\pi x)}=\frac{2\pi^2}{4^{x}\sin(\pi x)}.
\end{equation}
Putting all pre-factors in, one obtains the divergent behavior of the bosonic enhancement factor as: 
\begin{equation}
\label{Sk}
    S(\kappa\rightarrow 0)\approx\frac{2\pi^{3/2}\kappa^{2x-2}}{4^{x}\sin(\pi x)\Gamma(x+1/2)\zeta(x+1)}.
\end{equation}

For $0<x<1$, the divergence in Eq.~\ref{Sk} can be understood in an intuitive way. In $d$ spatial dimensions, the first-order correlation function of the non-interacting Bose gas approaches $g^{(1)}(r)\sim1/r^{d-2}$ at the critical point. Because the structure factor of a homogeneous sample $S_\text{box}(k)$ is related to the Fourier transform of the pair correlation function $g^{(2)}(r)-1=g^{(1)}(r)^2\sim1/r^{2d-4}$, $S_\text{box}(k)$ diverges as $\kappa^{d-4}$ when $\kappa \rightarrow0$. For a Bose gas trapped in a power-law potential $V(r)=V_0 (r/L)^\alpha$ ($L$ is the characteristic length of the system), only a fraction of the gas is close to local criticality at $T=T_c$. This fraction can be estimated from the characteristic length scale $r_0$ where the trapping potential energy is similar to the recoil energy: $V_0(r_0/L)^\alpha\sim k^2/2m$. Therefore, the fraction $f\sim(r_0/L)^d\sim \kappa^{2d/\alpha}$. Qualitatively, we can treat the sample within $r_0$ as homogeneous and apply the result for $S_\text{box}(k)$ at criticality to this small region. In the end, we get the structure factor for the trapped gas as $S(k)\sim \kappa^{2d/\alpha}\kappa^{d-4}\sim \kappa^{2x-2}$. 

For $x>1$, the bosonic enhancement mainly comes from regions outside the critical region which is very small, therefore the overall enhancement factor is finite. 

\section{Interaction effects on light scattering}
\textbf{Density distribution} 
The repulsive interaction between atoms modifies the density distribution in the trapping potential and greatly reduces the overlap between the thermal cloud and the condensate. The density distribution of the cloud was obtained using the local density approximation. Locally, the Bose gas is treated as a homogeneous gas with a Hartree-Fock (HF) mean-field Hamiltonian~\cite{pethick}:

\begin{equation}
  H_\text{HF} - \mu N = -gnN - \frac{g n_0 N_0}{2} + \sum_{k} \left(\epsilon_k - \mu + 2gn\right)a^\dagger_k a_k.
\end{equation}
 Here $n_0$ represents the density of the BEC while $n$ is the total density. $\mu$ is the chemical potential, and $g=4\pi\hbar^2a/m$ represents the interaction strength. $N$ and $N_0$ are the total atom number and atom number in the condensate, respectively. $a_k^\dagger$ and $a_k$ are the creation and annihilation operators for a free particle with momentum $k$ and kinetic energy $\epsilon_k=k^2/2m$. For a given chemical potential, the local density is obtained by solving the following pair of equations self-consistently:
\begin{equation}
  \left\{
  \begin{aligned}
    n &= n_0 + \frac{1}{\lambda_t^3}\, g_{3/2}(e^{\beta(\mu - 2gn)}) \\
    \mu &= 2gn - gn_0 \\
  \end{aligned}\right.
\end{equation}

\begin{figure}
\includegraphics[width=0.5\columnwidth]{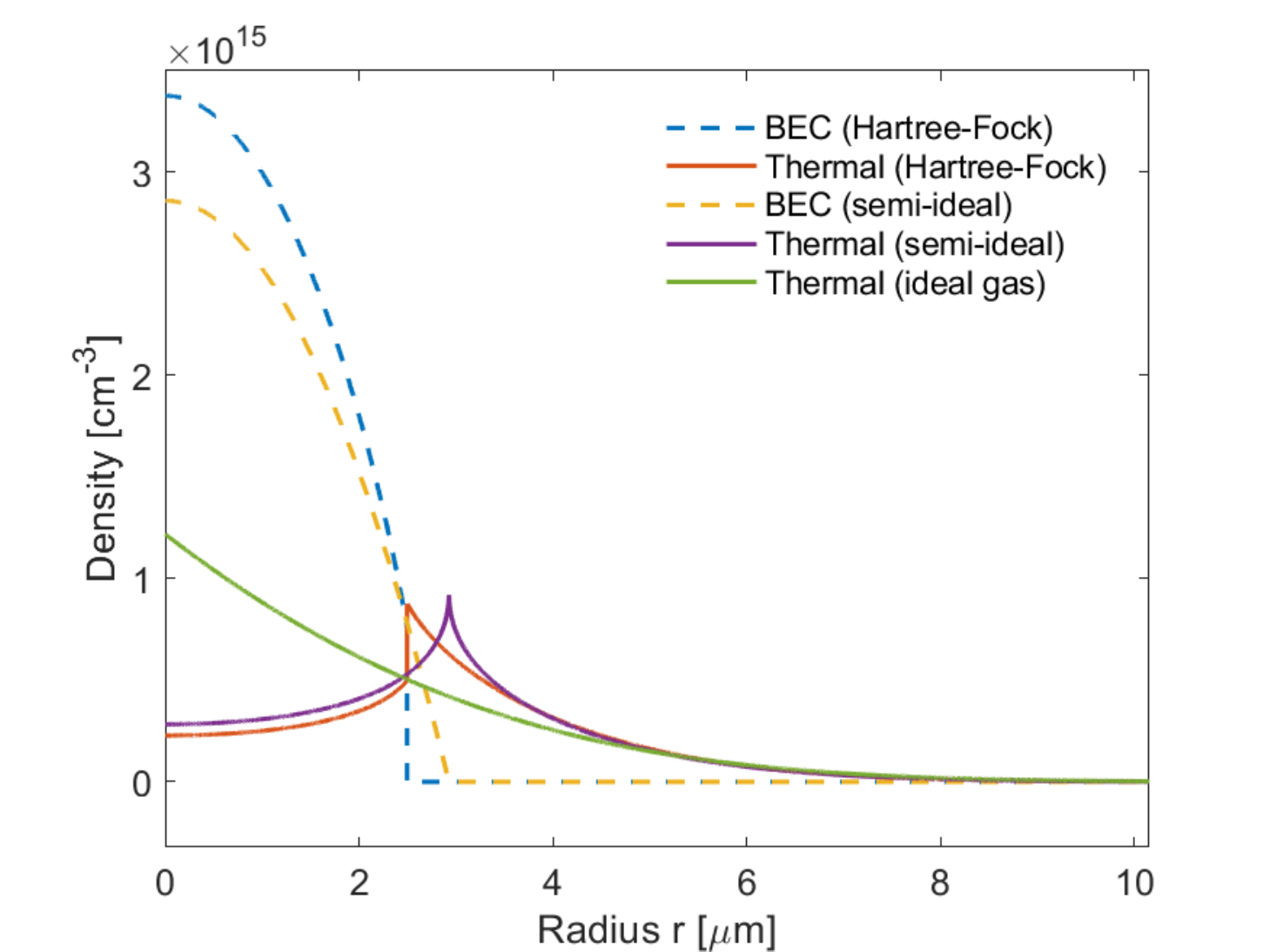}
\caption{\label{density} \textbf{The density distribution of the condensate and thermal cloud for different models.} In the calculations we used the following values: scattering length of $a=85a_0$ ($a_0$ is the Bohr radius), trap frequency $\omega=2\pi\times2.7\text{kHz}$, atom number $N=4\times 10^5$ and condensate fraction of 30\%.}
\end{figure}

We used an open-source package to solve these equations numerically for a harmonic trapping potential~\cite{hfsolver}. Note that here all calculations were done by assuming an isotropic harmonic trap with the mean trap frequency ${\omega}=(\omega_x\omega_y\omega_z)^{1/3}$. Unlike our previous work with fermions~\cite{SMPB}, the anharmonicity corrections for bosons here are much smaller and will be neglected. Fig.~\ref{density} shows the density profiles of a Bose gas for an ideal gas, in the semi-ideal gas approximation and HF approximation. Ref.~\cite{PhysRevA.81.053632} discusses the differences between the three models and compares them to experimental data. The ideal gas approximation assumes no interaction between atoms. The semi-ideal gas approximation considers the interaction within the BEC and between BEC and the thermal cloud, but not the interaction within the thermal cloud and the backaction of the thermal cloud onto the condensate. The mean-field repulsion of the thermal cloud by the condensate significantly reduces their overlap, causing the reduction of Bose enhancement. As a comparison, the HF approximation takes all interactions into account at the mean-field level. The calculations in Fig.~\ref{density} show that the backaction from the thermal cloud compresses the BEC to higher densities and further reduces their overlap by around 20\%. However, the HF calculation predicts an unphysical jump of the BEC density (Fig.~\ref{density}). For modeling the density distribution in the presence of a condensate, we used the semi-ideal gas approximation since it is numerically a much simpler approach. For a fully consistent calculation of the density profiles, one would have to go beyond the HF approximation.

To understand the importance of the different interaction terms, it is useful to compare the peak density of the thermal cloud at the phase transition ($n_\text{th}$) to the peak density of the BEC at $T=0$ ($n_{0}$). By using $n_\text{th}\propto N/(k_BT_c/m\omega^2)^{3/2}$ and $k_BT_c=\hbar \omega (N/\zeta(3))^{1/3}$, we get $n_\text{th}\propto N^{1/2}/a_\text{ho}^3$ ($a_\text{ho}=\sqrt{\hbar/m\omega}$ is the oscillator length). For the condensate, we can use $n_0=\mu_0/g$ with $\mu_0=\hbar\omega(15Na/a_\text{ho})^{2/5}/2$ and $g=4\pi\hbar^2a/m$ to obtain $n_0\propto N^{2/5}/(a^{3/5}a_\text{ho}^{12/5})$. Thus, the ratio of the densities of the thermal cloud and the BEC becomes $n_\text{th}/n_{0}\propto N^{1/6}a/a_\text{ho}$. Although the conventional wisdom is that the density of the BEC is usually much higher than the thermal cloud, $n_\text{th}$ and $n_0$ differ by only a factor of two in our experiment because of the tight confinement (Fig.~\ref{density}). Since the mean-field interaction of the thermal cloud has an extra factor of two due to the exchange term, we have a situation where the interactions in the condensate and the thermal cloud are equally important.

\textbf{Pair correlation function}
Atomic interactions will affect light scattering also through the pair correlation function. The exact expression of the pair correlation function for the non-interacting Bose gas reads~\cite{SMPhysRevA.59.4595}:

\begin{equation}
    G^{(1)}(r)=\sum_{l=1}^\infty\frac{e^{l\beta\mu}}{\lambda_t^3l^{3/2}}e^{\frac{-\pi r^2}{l\lambda_t^2}}.
\end{equation}

At high temperatures, the $l=1$ term dominates, so the correlation function is a Gaussian with a correlation length $\lambda_t/\sqrt{2\pi}$. Close to the critical point, the correlation function decays exponentially with a correlation length $\xi$~\cite{proukakis2017universal}:
\begin{equation}
    g^{(1)}{(r)} \sim \frac{\lambda_{t}}{\zeta(3 / 2)r} e^{-r/\xi}, \quad \xi=\frac{1}{\sqrt{-2 m \mu}}=\frac{\lambda_{t}}{\sqrt{-4 \pi\beta \mu}}.
\end{equation}
Here $g^{(1)}(r)=G^{(1)}(r)/G^{(1)}(0)$ is the normalized first-order correlation function. The correlation length $\xi$ diverges as $\left|T-T_c\right|^{-1}$ near the critical point for the non-interacting Bose gas.
Using the correlation function for the non-interacting gas, we can calculate the structure factor $S(k)$ for the ideal gas and the first-order correction of the structure factor with interactions~\cite{SMPhysRevA.59.4595}, $S_\text{int}(k)$. Near the critical point, we obtain:

\begin{equation}
    \begin{array}{l}
S(k=0)=\int_{0}^{\infty}\left|g^{(1)}{(r)}\right|^{2} 4\pi r^{2}  d r\propto\lambda_{t}^2\int_{0}^{\infty} \frac{e^{-2 r / \xi}}{r^{2}} 4 \pi r^{2} d r=2 \pi \lambda_{t}^2 \xi, \\
S_\text{int}(k=0)\approx S(k=0) -\int_{\lambda_{t}}^{\infty}\left|g^{(1)}{(r)}\right|^{2} \frac{4a}{r} 4\pi r^{2}  d r\propto2\pi\lambda_{t}^2\xi-16 \pi\lambda_{t}^2 a \Gamma\left(0, \frac{2\lambda_t}{\xi}\right) \sim 2\pi\lambda_{t}^2\xi-16 \pi\lambda_{t}^2 a \ln \frac{\xi}{2\lambda_{t}}.\\
\end{array}
\end{equation}

This expression shows that the relative correction due to interactions $S_\text{int}/S-1\sim-8a\ln(\xi/\lambda_t)/\xi$ vanishes as we approach the critical point. Note that the calculations were done by assuming $k=0$ because the correction for finite momentum is quadratic in $k$ and can be neglected.

At high temperatures, $g^{(1)}(r)\propto e^{{-\pi r^2}/{\lambda_t^2}}$ and the structure factor and its interacting correction are given by~\cite{SMPhysRevA.59.4595}:
\begin{equation}
\label{highTcorr}
    \begin{array}{l}
S(k=0)=\int_{0}^{\infty}\left|g^{(1)}{(r)}\right|^{2} 4\pi r^{2}  d r\propto\int_{0}^{\infty} e^{{-2\pi r^2}/{\lambda_t^2}} 4 \pi r^{2} d r=\frac{\lambda_{t}^3}{2\sqrt{2}}, \\
S_\text{int}(k=0)= S(k=0) -\int_{0}^{\infty}\left|g^{(1)}{(r)}\right|^{2} \frac{4a}{r} 4\pi r^{2}  d r\propto\frac{\lambda_{t}^3}{2\sqrt{2}}\left(1-8\sqrt{2}\frac{a}{\lambda_t}\right). \\
\end{array}
\end{equation}
Therefore, the correction factor for interaction effects approaches $1-8\sqrt{2}{a}/{\lambda_t}$ at high temperatures.
Note that the interacting pair correlation function describes the true correlation in the position of the atoms, whereas the non-interacting second-order correlation function $g^{(2)}_0(r=0)$ which is 1 for a pure condensate and 2 for a thermal cloud appears in the interaction parameters for the condensate and the thermal cloud, respectively~\cite{PhysRevA.56.3291, SMPhysRevA.59.4595}.

\begin{figure}
\includegraphics[width=1\columnwidth]{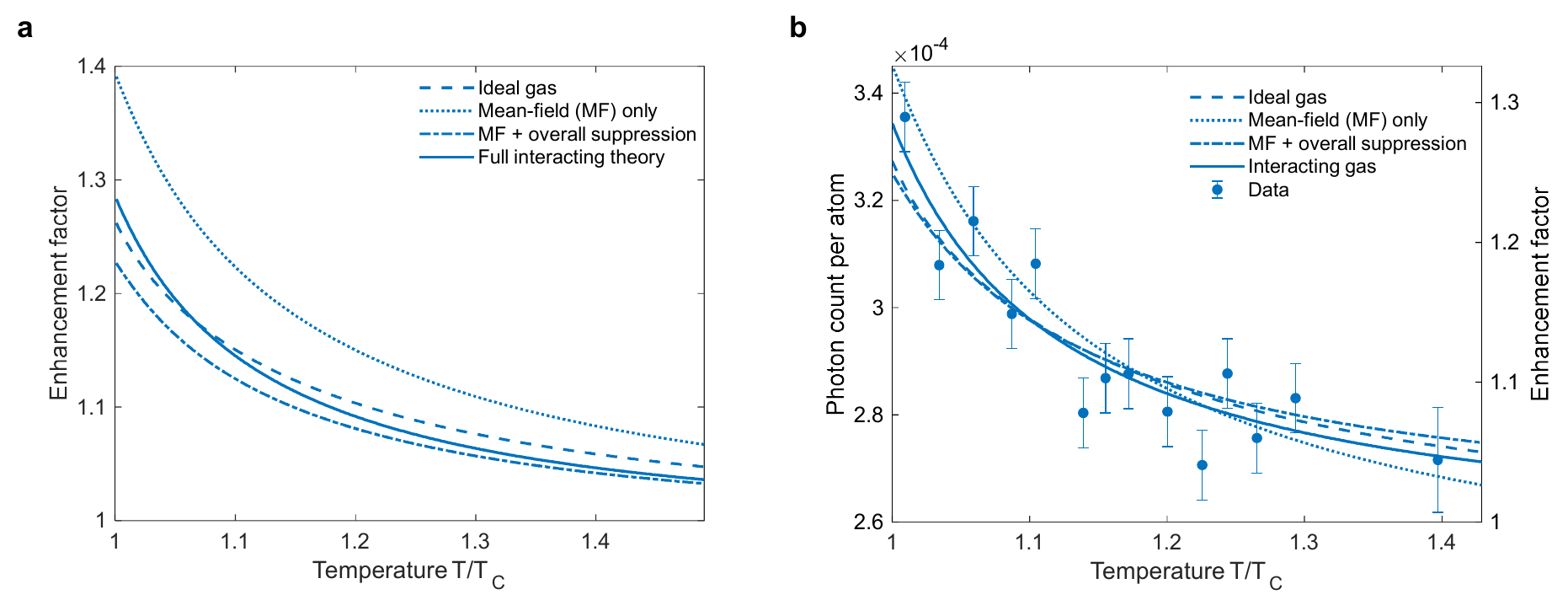}
  \caption{\label{TheoryComp} \textbf{Comparison between different models above the phase transition.} (a) Predictions for the Bose enhancement factor for different models. (b) Fitting of the models to the experimental data. The only free parameter is the overall scaling. In the calculations we used the following parameters: scattering length $a=85a_0$ ($a_0$ is the Bohr radius), atom number $N=4\times 10^5$ and dimensionless recoil momentum $\kappa=0.51$. }
\end{figure}

\textbf{Comparing interacting models to experimental data above the phase transition}
The bosonic enhancement of light scattering is about 30\% just above the phase transition temperature. The mean-field modification of the density distribution increases the light scattering by about 10\%, whereas the modification of the pair correlation function leads to a decrease by about 10\%. This is why the predictions for the full theory for interactions are close to the ideal gas prediction. This is shown in Fig.~\ref{TheoryComp} where three different models are compared to the ideal gas: The first mean-field (MF) model includes only the modified density distribution of the cloud, but not the pair correlation function, while the second model (MF + overall suppression) approximates the effect of the modified pair correlation as an overall reduction factor of $1-8\sqrt{2}a/\lambda_t$. The third model (full interacting theory) includes the dependence of the pair correlation function on the local chemical potential. So we locally calculated the pair correlation function (without assuming a Gaussian or exponential form) and the structure factor $S_\text{int}(k)$ which gives the local scattering rate. The total light scattering rate is obtained by integrating the local rates across the cloud.

We fit the data using these three different models by only adjusting the overall scaling. A comparison of the $\chi^2$ values for the fits shows that the best fit of the data is provided by the full interacting model (table~\ref{TBchi2}). Therefore, our observations provide some evidence for the modification of pair correlations by interactions, and even for the different effects of the pair correlations near criticality. However, the statistical significance is modest, as illustrated in Fig.~\ref{TheoryComp}b. Note that some of the differences in Fig.~\ref{TheoryComp}a disappeared because the high-temperature asymptotic value is a fit parameter in Fig.~\ref{TheoryComp}b. Unfortunately, we couldn't measure the asymptotic value at higher temperatures because of the limitation by the trap depth. We suggest that future experimental studies use a box potential instead of a harmonic potential to avoid the partial cancellation of mean-field and pair correlation effects.

\textbf{Details of the interacting model below the phase transition} The full interacting theory below the phase transition becomes very complex, and we did not aim for a fully quantitative description. Therefore, we used the following approximations for the theory curves in Fig.~\ref{fig3}b: (1) the density distribution was obtained using the semi-ideal gas approximation. (2) The effect of the pair correlation function was introduced by a reduction factor of $1-8\sqrt{2}{a}/{\lambda_t}$ (Eq.~\ref{highTcorr}). Note that the theory curves in Fig.~\ref{fig3}b has no free parameters (the overall scaling was the same as Fig.~\ref{fig3}a).

\begin{table}
\begin{tabular}{|c|c|c|c|c|}
\hline
                  & Ideal gas & Mean-field (MF) only & MF + overall suppression & Full interacting theory \\ \hline
$\chi^2$ value    & 20.0      & 18.5                 & 21.4                     & 16.1                    \\ \hline
Normalized $\chi^2$ value  & 16.1       & 14.9                   & 17.3                       & 13                      \\ \hline
Degrees of freedom (dof) & 13        & 13                   & 13                       & 13                      \\ \hline
\end{tabular}%
\caption{\label{TBchi2}
\textbf{$\chi^2$ values for fits using different models.} 
The $\chi^2$ values are normalized in such a way that it becomes $\text{dof}=13$ for the full interacting model. 
As a comparison, the probability for $\chi^2$ (dof=13) to be larger than 15.1 is 30\% while the probability for $\chi^2$ to be larger than 17.0 is 20\%. 
}
\end{table}

%


\begin{thebibliography}{27}
\bibitem [{foo}({\natexlab{a}})]{footnote}%
  \BibitemOpen
  \href@noop {} {\bibinfo {title} {{Reference 15 claimed the observation of
  bosonic enhancement of light scattering. However, their observable was atom
  loss caused by optical pumping. As we show in the last part of the paper,
  optical pumping (i.e. Raman scatteing) into an unoccupied state does not show
  any bosonic enhancement. The larger observed loss is possibly caused by the
  higher atomic density in a Bose-Einstein condensate.}}}\BibitemShut
  {Stop}%
\end{thebibliography}
\end{document}